\documentstyle[sprocl,epsfig]{article}

\def\NPB{{\em Nucl. Phys.} B}
\def\PLB{{\em Phys. Lett.}  B}

\def\ZPC{{\em Z. Phys.} C}

\def\be{\begin{equation}}
\def\ee{\end{equation}}
\def\bea{\begin{eqnarray}}
\def\eea{\end{eqnarray}}

\newcommand{\beq}{\begin{equation}}
\newcommand{\eeq}{\end{equation}}
\newcommand{\beqn}{\begin{eqnarray}}
\newcommand{\beqqn}{\begin{eqnarray*}}
\newcommand{\eeqn}{\end{eqnarray}}
\newcommand{\eeqqn}{\end{eqnarray*}}
\newcommand{\refeq}[1]{(\ref{#1})}

\begin{document}

\title{TARGET FRAGMENTATION OF THE NUCLEON AT HIGH ENERGIES
}

\author{U. D'ALESIO
\footnote{Supported by TMR programme,
EU-Project FMRX-CT96-0008.}}

\address{Institut f\"ur Theoretische Physik der Universit\"at
Heidelberg,\\
69120 Heidelberg, Germany\\E-mail:
umberto@tphys.uni-heidelberg.de}


\maketitle\abstracts{ 
We calculate target fragmentation in $p p \rightarrow n X$
and $\gamma p \rightarrow n X $ reactions in the meson cloud
picture of the nucleon. 
The $p p \rightarrow n X$ reaction is used to fix the
$pn\pi^+$ form factor for two different models.
We take into account the possible destruction of
the residual neutron by the projectile. Using the form factor
from the hadronic reaction we calculate photoproduction and small
$x_{\rm Bj}$ electroproduction of forward neutrons at HERA. 
In photoproduction we observe slightly less  absorption
than in the hadronic reaction.
For deep inelastic events ($Q^2>10$ GeV$^2$) screening
is weaker but still present at large $Q^2$.
The signature for this absorptive rescattering
is a shift of the  $d\sigma/dE_n$ distribution to higher neutron
energies for photofragmentation.}

\section{Introduction}

In the one pion exchange model
the differential cross section for a 
general reaction 
leading to neutron production i.e. 
$ a p \rightarrow n(z,p_t) X $   reads\footnote{Hereafter $z$ is the neutron
longitudinal momentum fraction, $p_t$ its transverse momentum 
and $t$ the momentum transfer.}
\beq
\label{opeap}
E_n\frac{d^3\sigma}{d^3 {\bf p}_n} = 
   	 \frac{2g^2}{16\pi^2}\frac{|t|}{(t-m_\pi^2)^2}
   (1-z)^{1-2\alpha_\pi(t)}
         |G(z,p_t)|^2 \sigma_{\rm tot}^{a\pi }\,,
\eeq
where the first factor is the splitting probability of a proton into a
neutron-pion system (including the reggeization of
the pion in a covariant approach), 
$G(z,p_t) = exp[-R^2_c(t-m_\pi^2)]$ is the $pn\pi$ form factor, and
$\sigma_{\rm tot}^{a\pi }$ is the
total $a\pi$ cross section.
To get the lightcone formula  one has to set $\alpha_\pi(t) = 0$ and 
to replace $R^2_c$ by $R^2_{lc}/(1-z)$.

The usual procedure to extract the pion structure function 
\cite{HLNSS,KPP,Ze,H1} is to fix the relevant
parameters entering the flux factor
from the data of leading neutron production in proton-proton ($a=p$)
collisions and  then apply eq.\refeq{opeap}
for virtual photon scattering ($a=\gamma^*$) in DIS.

Recently a first study of absorptive corrections in the Regge formalism
\cite {NSZ} has appeared. These absorptive effects
depend on the projectile and are a source of factorization
breaking.
Applying high-energy Glauber theory, we will investigate 
the relevance of
absorptive corrections in detail in order to  understand 
the one pion exchange mechanism and the extraction of the pion structure
function. 

\section{Estimate of absorptive 
corrections in $pp\rightarrow nX$}

We consider the target fragmentation reaction as a stripping 
reaction where the projectile proton
strips a $\pi^+$ from the 
the target proton, leaving behind a neutron.
This picture is reliable when the pion and the neutron in the {\it
target\/}
proton are well separated (as in the case of large $z$) but  
becomes questionable at intermediate values of $z$
when  
the rescattering of the projectile on the neutron and
its subsequent screening  can be important. 
We treat the target proton as a
pion-neutron system ($\phi_0$) undergoing a  
transition to an excited state
($\phi_\alpha$) and  sum over all excited states, excluding the
elastic contribution. The differential cross section reads 
\beq
\frac{d\sigma}{dz} = \int\!d^2b\int\! d^2b_{rel}
\rho_{\pi n}(z,{\bf b}_{rel})
2{\em Re}\Gamma_{p\pi }({\bf b - s}_{\pi})
[ 1-2{\em Re}\Gamma_{pn}({\bf b - s}_n)]\,,
\eeq
where ${\bf b}$ is the impact parameter, ${\bf s}_{\pi} = -z{\bf b}_{rel}$
and ${\bf s}_{n}= (1-z){\bf b}_{rel}$ are the coordinates\footnote{These
last relations come from the center-of-mass constraint.} 
of the pion and the neutron in the impact parameter plane,
and ${\bf b}_{rel}$ is the relative distance between the pion and the
neutron. For the profile functions 
we use $\Gamma_{ab}({\bf b}) = \sigma_{tot}^{ab}/(4\pi)\Lambda_{ab}^2
exp[-b^2\Lambda_{ab}^2/2]$.
The density factor $\rho_{\pi n}(z,{\bf b}_{rel})$ is the probability
to find the pion and the neutron at relative distance ${\bf b}_{rel}$ and
fixed momentum fraction $z$.

\begin{figure}[!htb]
\center
\mbox{\epsfig{file=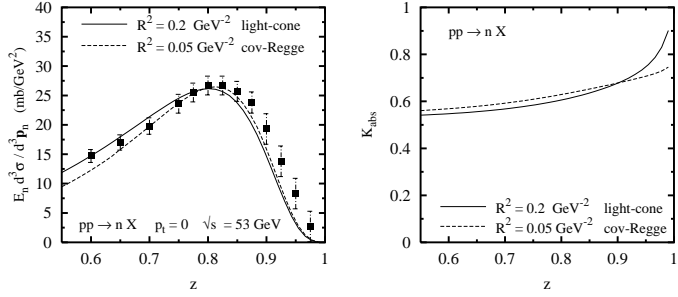,width=.75\textwidth}}
\caption{Differential cross sections  
at $p_t = 0$ and absorptive $K$-factors.}
\label{brus1}
\end{figure}

We calculate the invariant differential cross 
section  assuming  that the
final state interaction does not modify the transverse 
momentum distribution of the fragments.
We adjust the radius parameters $R_{lc}$  and $R_c$
in the light-cone and in the covariant form factors respectively to the
experimental data at $p_t=0$ \cite{FlMo}. The additional background is
included rescaling the pion exchange contribution by a factor 1.2. 
We find  reasonable agreement with the data for a
radius squared $R_{lc}^2$=  0.2 GeV$^{-2}$  and 
$R_{c}^2$ = 0.05 GeV$^{-2}$, (see fig.~\ref{brus1}). 
The screening effect  can be clearly seen in the 
same figure, where we plot the $K$-factor, i.e. the 
ratio of the differential cross section with and without absorptive
corrections.

\section {Cross sections for $\gamma^* p \rightarrow n X$ and 
$\gamma p\rightarrow n X$}

We limit ourselves to forward neutrons in 
photon induced reactions at
small  $x_{\rm Bj}$, as studied at HERA. 
For small $x_{\rm Bj}$, we  consider the
photon as a quark-antiquark state which materializes long 
before the proton and interacts with the 
pion and neutron in the proton wave function. 
Schematically, we get the inelastic cross section in a form similar 
to the proton induced cross section
\beqn
\label{gpabs}
\frac{d\sigma}{dz} & = & 
        \int\!d^2{b}_{rel}\rho_{n\pi}(z,{\bf b}_{rel})\,
        \int\! dw d^2{r}|\Psi_{q \bar q }(w,{\bf r})|^2
      \sigma_{\rm tot}^{q\bar q\pi}({\bf r})     \nonumber\\
& &   \Big\{1-\Lambda_{\rm eff}^2\frac{\sigma_{\rm tot}^{q
        \bar qn}({\bf r})}{2\pi}\,
        {\rm exp}\Big[-\frac{\Lambda_{\rm eff}^2{\bf
        b}_{rel}^2}{2}\Big]\Big\}\,, \;\;\;
\Lambda_{\rm eff}^2 = \frac{\Lambda_{q\bar q\pi}^2\Lambda_{q\bar q
        n}^2}{\Lambda_{q\bar q\pi}^2+\Lambda_{q\bar q n}^2}.
\eeqn
Here the $q \bar q$ pair wave function is represented by 
$|\Psi_{q \bar q }(w,{\bf r})|^2$ with $w$ as the momentum fraction of
the
quark and $\bf r$ as the $q\bar q$ transverse separation.
Note that it  does not enter in the magnitude
of the screening correction with the same weight as in the direct term. 
Screening is  a strong
interaction effect which is a function of ${\bf r}$.
Eq.~\ref{gpabs} can be recast as
\beq
\label{shadeff}
\frac{d\sigma}{dz} = \int\!d^2{b}_{rel}
        \rho_{n\pi}(z,{\bf b}_{rel})\,\sigma_{\rm tot}^{\gamma^*\pi^{+}}
        \Big\{1-\Lambda_{\rm eff}^2
        \frac{\sigma_{\rm eff}}{2\pi}\,
        {\rm exp}\Big[-\frac{\Lambda_{\rm eff}^2{\bf
        b}_{rel}^2}{2}\Big]\Big\}\,,
\eeq
\beq
\label{sigratioKo}
{\rm with\/}\;\;\sigma_{\rm eff} \equiv \langle\sigma_{\rm tot}^{q\bar q\pi}
        \sigma_{\rm tot}^{q\bar qn}\rangle/
        \langle\sigma_{\rm tot}^{q\bar q\pi}\rangle = 
        N_0\frac{1}{F_2^p(x_\pi)}
        \left(\frac{1}{x_\pi}\right)^{\Delta_{\rm eff}}
        \left(\frac{1}{x_n}\right)^{\Delta_{\rm eff}} \,.
\eeq
where we have included the proper scaling variable dependence.
We use $N_0 = 2$ GeV$^{-2}$ and $\Delta_{\rm eff} = 0.15$.
The low value for $\Delta_{\rm eff}$ comes from the fact 
that diffraction and therefore shadowing
are dominated by soft processes. For real photons we use 
$ \sigma_{\rm eff}|_{Q^2=0} \approx  20\, {\rm mb}$.
\begin{figure}[!htb]
\center
\mbox{\epsfig{file=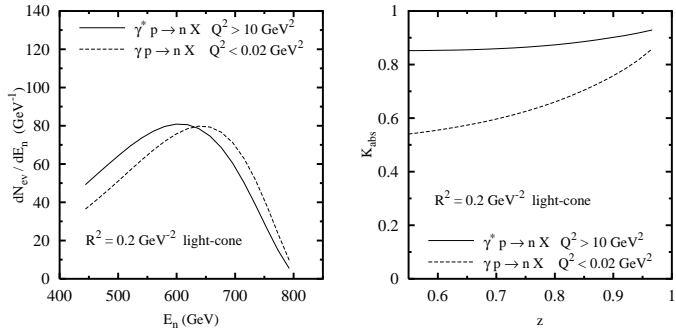,width=.75\textwidth}} 
\caption{\label{brus2} 
 Energy distributions of neutrons for deep inelastic
events and photoproduction normalized to the total number of DIS events
(left) and absorptive $K$-factors (right).
}
\end{figure}
In fig.~\ref{brus2}  we plot the integrated photoproduction 
 and deep inelastic energy distributions 
applying the ZEUS cuts \cite{Ze}. 
A shift of the peak to a 50 GeV lower energy
in the $dN/dE_n$ for DIS  neutron production is
visible. 
This comes from the effective screening
in the photoproduction, which eats up cross section at 
smaller $z$, making the peak appear at higher energies. 
At large $Q^2$ screening is reduced but still non negligible. 
The  $K$-factors for $\gamma p\rightarrow 
nX$ and $ep\rightarrow e'nX$ are shown in fig.~\ref{brus2}.

\section{Discussion of fragmentation results and 
validity of the factorization hypothesis}

Finally, we compare the screening corrections 
for the three different cases of proton, real and DIS photon induced
semi-inclusive fragmentation reactions. 
Both proton induced and real
photon induced cross sections have $K$-factors differing by about 
30\% from unity for $z<0.85$, while 
for highly virtual photons this effect is reduced.
This makes a model-independent extraction 
of the pion structure function difficult for these $z$-values. 
Thus semi-inclusive neutron fragmentation,
even on the nucleon, seems to be a new channel available for the study of
final-state interactions of virtual partons \cite{DPS}.



\end{document}